\documentclass[aps,prd,onecolumn,groupedaddress,showpacs,nofootinbib,amssymb]{revtex4}
\usepackage[dvips]{graphicx}
\usepackage{amssymb}
\usepackage{amsmath}
\usepackage{graphicx,,color}
\usepackage{amsfonts}
\usepackage{bm}
\usepackage{xcolor}
\usepackage{cancel}
\usepackage{comment}
\newcommand\be{\begin{equation}}
\newcommand\ee{\end{equation}}

\allowdisplaybreaks[4]

\begin{document}
\tolerance=5000

\title{Inflation with exotic kinetic terms in Einstein-Chern-Simons gravity}
\author{F.P.
Fronimos,\, \thanks{fotisfronimos@gmail.com}S.A.
Venikoudis\,\thanks{venikoudis@gmail.com}}
\affiliation{
 Department of Physics, Aristotle University of
Thessaloniki, Thessaloniki 54124,
Greece\\}

\tolerance=5000


\begin{abstract}
An alternative scenario about the phenomenology of primordial Universe is  
k-inflation. According to this concept, inflation can be achieved by nonstandard kinetic term of scalar field, namely the inflaton. In this project we focus on k-essence models in the presence of a higher order and a linear kinetic term. Furthermore, the inflationary phenomenology with a Dirac-Born-Infeld scalar field is briefly examined, which arises from quantum theories of gravity such as superstring theory. Our approach about the inflationary era is that it can be described in the context of Einstein's gravity involving quantum corrections such as the Chern-Simons string inspired parity violating gravitational term.
The equations of motion namely, the Friedmann equation, the Raychadhuri equation and the Klein-Gordon equation for an expanding background are extracted from the gravitational action utilizing the variational principle. The consequential system of differential equations with respect to Hubble's parameter and the inflaton field was quite perplexed in order to be solved with an analytic way. Therefore, the slow-roll conditions during inflationary era were imposed and terms with minor numerically contribution were neglected. From the overall phenomenological analysis it is proved that, models with exotic kinetic terms can generate viable results in consistency with the latest Planck data. Finally, the presence of Chern-Simons quantum corrections shifts the primordial spectral tensor index to blue. Even though blue gravitational waves have yet to be observed, if detected, compatibility with the aforementioned theory can be achieved.

\end{abstract}
\maketitle

\section{Introduction}
One of the most fascinating mysteries in modern Cosmology is the phenomenology of primordial Universe. 
The inflationary paradigm provides the most effective way to describe the dynamics of primordial era of our Universe, solving many discrepancies of standard Big Bang cosmological model such as the horizon and flatness problem, the initial entropy problem and the explanation for the non-existence of magnetic monopoles.
According to inflationary mechanism the early Universe expanded with an accelerating rate. In the simplest case, a scalar field, corresponding to the quasi de-Sitter expansion, slowly rolls down the potential \cite{Guth:1980zm}.

An alternative approach of this theory is the kinetically driven inflation(k-inflation),
which was first introduced by Mukhanov, Damour and Armendariz-Picon \cite{Armendariz-Picon:1999hyi}. According to this pioneer work, the inflaton may involve higher-order kinetic terms, for instance non-quadratic. Furthermore, accelerating expansion can be achieved with or without the existence of scalar potential. Even though the observational constraints of inflation become more accurate the recent years, the effective Lagrangian of inflation has not been specified by the data until today. Based on the effective field theory of inflation \cite{Cheung:2007st}, the Primordial era of our Universe can be described in the context of a generalized way in classical Einstein's gravity with the presence of quantum fluctuations around a quasi de Sitter background.
According to this reasoning, inflationary era dates in between the classical era and the
quantum era where all fundamental forces were unified and the Universe was governed by a quantum theory of gravity. It is reasonable to consider that the effective Lagrangian may involves higher gravitational terms like $f(R)$ gravitational terms \cite{Nojiri:2017ncd,Capozziello:2011et,Capozziello:2010zz,Nojiri:2006ri,Nojiri:2010wj,Olmo:2011uz,Venikoudis:2021irr} and also Chern-Simons terms \cite{Venikoudis:2021oee,Nojiri:2020pqr,Nojiri:2019nar,Odintsov:2019evb,Odintsov:2019mlf,Alexander:2009tp,Qiao:2019hkz,Nishizawa:2018srh,Wagle:2018tyk,Yagi:2012vf,Yagi:2012ya,Molina:2010fb,Izaurieta:2009hz,Smith:2007jm,Konno:2009kg,Sopuerta:2009iy,Matschull:1999he,Haghani:2017yjk}. In this work we focused on the impact of Chern-Simons corrections on the inflationary phenomenology and the primordial gravitational waves which can be produced. Specifically, the existence of Chern-Simons term in Lagrangian does not affect the background equations and the scalar spectral index of primordial curvature perturbations, but it has strong impact on tensor perturbations of a given theory.
We shall also investigate whether is possible the tensor spectral index to be shifted to the blue \cite{Camerini:2008mj}. Primordial gravitational waves with blue tensor spectral index have not been detected yet, but if this happen, then the existence of Chern-Simons term in the effective Lagrangian of inflation might be involved. In addition, based on Huang and Noh paper \cite{Hwang:2005hb} the Chern-Simons parity violating term leads to asymmetric generation and evolution of two circular polarization states of gravitational waves.

K-essence models provide  an attractive class of inflationary  models to be examined, as they generate results compatible with observational constraints
see, Ref.\cite{Nojiri:2019dqc,Oikonomou:2019muq,Mohammadi:2019qeu,Myrzakul:2016ets,Sebastiani:2016ojm,Rahmati:2014cwa,Saitou:2011hv,Nojiri:2010aj,Arroja:2010wy,Bose:2008ew}.
In this project it is assumed that the k-essence scalar field is minimally coupled with the scalar curvature and involves a higher order and a linear kinetic term. Firstly, the theoretical framework of k-essence overall phenomenology is presented for the purpose of examining the impact of Chern-Simons term in dynamics of inflation. Furthermore, the observational indices are presented in order to ascertain the viability of models. The first proposed model involves a power-law scalar potential and an exponential scalar Chern-Simons coupling function while the second involves a trigonometric potential and a power-law scalar Chern-Simons coupling function. Both models admit a quadratic noncanonical kinetic term.

Apart from k-essence models, there is another class of scalar fields which produce viable kinetically driven inflation namely Dirac-Born-Infeld scalar field. These fields arise from quantum theories of gravity such as string theory. Based on string theory the relativistic motion of a D3 brane in a higher dimensional warped spacetime generates D-brane inflation with Dirac-Born-Infleld scalar kinetic term, see Ref.\cite{ Nguyen:2021emx,Yang:2021enz,Motaharfar:2021egj,Chakraborty:2020kjy,Chen:2020uhe,Rashidi:2020wwg,Nozari:2019esz,Seo:2018abc,Mohammadi:2018zkf,Lahiri:2018erp,Nazavari:2016yaa,Bielleman:2016grv,Nozari:2013wua,Zhang:2013asa,Koehn:2012np,Miranda:2012rm,Weller:2011ey,Chimento:2010un,Franche:2010yj,Bessada:2009pe,Arroja:2009pd,Langlois:2008wt,Bean:2007eh,Lidsey:2007gq,Spalinski:2007dv,Spalinski:2007qy,Spalinski:2007un,Kecskemeti:2006cg,Chen:2005fe}. In this paper we shall examine the phenomenology considering that during inflationary era the scalar potential remains constant for simplicity.

\section{INFLATIONARY PHENOMENOLOGY OF K-ESSENCE MODELS IN EINSTEIN-CHERN-SIMONS GRAVITY}
The k-essence inflationary models can be described in the context of $f(R,\phi,X)$ class of theories where, R denotes the Ricci scalar, $\phi$ stands for the scalar field with X being $X=\frac{1}{2}\nabla^{\mu} \phi \nabla_{\mu}\phi$. The proposed gravitational action during the inflationary era includes the k-essence scalar field in the presence of Chern-Simons term and can be defined as,
\begin{equation}
\centering
\label{action}
S=\int {d^4x\sqrt{-g}\left[\frac{f(R,\phi,X)}{2}+ \frac{1}{8}\nu(\phi)R\Tilde{R}\right]},
\end{equation}
where $g$ is the determinant of the metric tensor and $\nu(\phi)$ is the Chern-Simons scalar coupling function. The Chern-Simons term represents parity violation in gravity and it is given by the expression  $R\Tilde{R}=\epsilon^{a bcd}R^{\ \ e f}_{a b}R_{cdef}$ where, $\epsilon^{abcd}$ is the totally antisymmetric Levi-Civita tensor in 4-dimensions. The line-element is assumed to have the
Friedmann-Robertson-Walker form,
\begin{equation}
\centering
\label{metric}
ds^2=-dt^2+a(t)^2\delta_{ij}dx^idx^j,\,
\end{equation}
where $a(t)$ is the scale factor of the Universe and the metric tensor has the form of $g_{\mu\nu}=diag(-1, a(t)^2, a(t)^2, a(t)^2)$. It is both sensible and convenient to assume that the scalar field participating in the aforementioned gravitational action is homogeneous, or in other words only time dependent. In consequence the kinetic term, which is of paramount importance in the k-essence model can be simplified as $X=-\frac{1}{2}\dot\phi^2$, where the dot as usual implies differentiation with respect to cosmic time $t$.
Additionally, the general form of the $f(R,\phi,X)$ function can be considered as,
\begin{equation}
\centering
\label{f}
f(R,\phi,X)=\frac{R}{\kappa^2}-2\alpha X^m-2\alpha_0 X-2V(\phi),
\end{equation}
where $\kappa=\frac{1}{M_P}$ is the gravitational constant while, $M_P$ denotes the reduced Planck mass, and $V(\phi)$ is the scalar potential. Note that in the present form, the auxiliary parameters introduced, in particular $\alpha_0$ and $\alpha$ must have proper dimensions. In particular, while the first is to be considered as dimensionless with possible values of $0,\pm1$, with $-1$ specifying phantom scalar field, the latter must have mass dimensions of eV$^{4-4m}$ for consistency. Hereafter, parameter $\alpha_0$ shall be assumed to be equal to unity, unless stated otherwise however it is presented in subsequent equations as free parameters for the sake of generality. In addition, parameter $m$ is to be replaced with $m=2$ in the following k-essence toy models however in order to be inclusive the general case for an arbitrary exponent $m$ is showcased. Finally, as long as the metric is flat, the Ricci scalar term is topological invariant and can be written as $R=12H^2+6\dot H$ where $H$ is Hubble's parameter and in addition, the ``dot'' denotes differentiation with respect to the cosmic time. Let us now proceed to the proper phenomenology of the inflationary era.

The equations of motion for the theory can be derived by implementing the variational principle in the gravitational action (\ref{action}) with respect to the metric tensor and the scalar field separately which read,
\begin{equation}
\centering
\label{motion1}
3H^2F=f_{,X}X+\frac{FR-f}{2}-3H\dot F,\,
\end{equation}
\begin{equation}
\centering
\label{motion2}
-2\dot H F=f_{,X}X+\Ddot{F}-H\dot F,\,
\end{equation}
\begin{equation}
\centering
\label{motion3}
\frac{1}{\alpha^3}\frac{d}{dt}(\alpha^3 \dot \phi f_{,X})+f_{,\phi}=0 \
\end{equation}
where $F=\frac{\partial f}{\partial R}$ represents the partial derivative with respect to the Ricci scalar while, $f_{,X}=\frac{\partial f}{\partial X}$, $f_{,XX}=\frac{\partial f_{,X}}{\partial X}$ and finally $f_{,\phi}=\frac{\partial f}{\partial\phi}$. The first describes the Friedmann equation or the temporal component of the field equations, the second being the Raychadhuri equation and the third stands for the continuity equation of the scalar field or in other words the Klein-Gordon equation for an expanding background. In essence, the Klein-Gordon equation for the aforementioned model is quite nontrivial due to the additional noncanonical kinetic term. Substituting Eq.$(\ref{f})$ into the equations of motion the resulting equations are,
\begin{equation}
\centering
\label{motion11}
\frac{3H^2}{\kappa^2}=-\alpha_0 X-\alpha X^m(2m-1)+V(\phi),\,
\end{equation}
\begin{equation}
\centering
\label{motion22}
\frac{\dot H }{\kappa^2}=\alpha_0X+\alpha m X^m,\,
\end{equation}
\begin{equation}
\centering
\label{motion33}
\Ddot{\phi}[\alpha_0+\alpha m(2m-1)X^{m-1}]+3H\dot \phi(\alpha_0+\alpha m X^{m-1})+V'=0\
\end{equation}
where the prime denotes differentiation with respect to the scalar field $\phi$. Obviously, the resulting system of equations is perplexed to be solved with an analytic way. On account of this, the slow-roll approximations are imposed as long as inflation last. Specifically, the following approximations are considered to hold true,
\begin{align}
\label{approx}
\centering
\dot H&\ll H^2,& \mid \alpha_0 X+\alpha m X^m \mid&\ll V,& \ddot\phi\ll3 H\dot\phi,\
\end{align}
therefore, the equations of motion can be simplified greatly. In the following sections it is proved numerically that indeed these conditions are in effect for the presented models. The simplified equations are, 
\begin{equation}
\centering
\label{motion11}
H^2\simeq\frac{\kappa^2V}{3},\,
\end{equation}
\begin{equation}
\centering
\label{motion33}
3H\dot \phi(\alpha_0+\alpha m X^{m-1})+V'\simeq 0,\
\end{equation}
while Eq.(\ref{motion22}) remains the same and carries the impact of the k-essence term. Considering $m=2$ for simplicity as expressed previously and substituting the kinetic term X into the continuity equation for the k-essence field, Eq.(\ref{motion33}) reduced to the simple form,
\begin{equation}
\centering
\label{scalar}
3H \dot \phi(\alpha_0-\alpha \dot \phi^2)+V' \simeq 0,
\end{equation}
where the quadratic term of the derivative of the scalar field with respect to cosmic time is negligible compared to $\dot \phi$. One can easily obtain the expression of the time derivative of the scalar field as,
\begin{equation}
\centering
\label{fdot}
\dot \phi \simeq -\frac{V'}{3H\alpha_0}.
\end{equation}
The dynamics of inflation in the context of $f(R, \phi, X)$ gravity is described in terms of the slow-roll parameters based on Huang and Noh paper \cite{Hwang:2005hb} as,
\begin{align}
\centering \epsilon_1&=-\frac{\dot
H}{H^2},&\epsilon_2&=\frac{\ddot\phi}{H\dot\phi},&\epsilon_3&=\frac{\dot E}{2HE},&\epsilon_4&=\sum_{L,R}\frac{\dot Q_t}{2HQ_t},
\end{align}
where summation over $L$ and $R$ implies summation over left and right handed polarization of gravitational waves.
During the first horizon crossing, the 
parameter $Q_{tCS}$ is defined as,
\begin{equation}
 Q_{t_{CS}}=F+2\lambda_l\dot \nu H.\ 
\end{equation}
 Such auxiliary functions are important for studying the behavior of the gravitational wave modes. The analysis has already been performed in \cite{Hwang:2005hb}.
Finally, the parameter E which is given by the following expression,
\begin{equation}
E=-\frac{F}{2 X}\left(X f,_X+2X^2 f,_{XX}\right),
\end{equation}
is reduced to the form,
\begin{equation}
\centering
E=\frac{1}{\kappa^2}\left(\alpha_0+\alpha m(2m-1)X^{m-1}\right).
\end{equation}
The slow-roll indices for unspecified scalar potential and Chern-Simons scalar coupling function are given by the following expressions,
\begin{equation}
\label{e1k}
\centering
\epsilon_1 \simeq  \frac{V'^{2}}{ \alpha_0 \kappa^2 V^2}(\frac{1}{2}-\frac{\alpha V'^{2} }{7\kappa^2 \alpha_0^3 V}) ,
\end{equation}
\begin{equation}
\label{e2k}
\centering
\epsilon_2\simeq \epsilon_1-\frac{V''}{\kappa^2 V},
\end{equation}
\begin{equation}
\label{e3k}
\centering
\epsilon_3\simeq \frac{\alpha V'^{2}(2VV''-V'^{2})}{2\kappa^2 V^2(\alpha_0^4 \kappa^2 V-\alpha_0 \alpha V'^{2})},
\end{equation}
\begin{equation}
\label{e4k}
\centering
\epsilon_4\simeq \frac{4\kappa^2 \nu' V'^{2}(V'\nu''+\nu'V'')}{\alpha_0 V(9\alpha_0^2-4\kappa^4 \nu'^{2}V'^{2})} .
\end{equation}
The involvement of Chern-Simons quantum corrections affects only the index $\epsilon_4$ and in consequence only the tensor observed indices presented in the following section.

\section{Compatibility of INFLATIONARY MODELS WITH THE LATEST PLACK DATA}
At this point it is worth to be mentioned that Cosmology is at a phase where the observational constraints needs to be explained theoretically. Thus, our goal is to speculate and fit the parameters of the aforementioned theory in order to produce results compatible with the latest Planck data \cite{Planck:2018jri} and consequently viable. In this section we propose models which are consistent with the observations. Firstly, in Einstein-Chern-Simons gravitational theory according to Ref. \cite{Hwang:2005hb} we shall evaluate the following observational quantities, namely the scalar spectral index of primordial curvature perturbations $n_\mathcal{S}$, the tensor-to-scalar-ratio $r$ and finally, the tensor spectral
index $n_\mathcal{T}$. The observational indices are strongly related with the slow-roll indices which described by the Eq.(\ref{e1k}-\ref{e4k}),
\begin{align}
\label{observed}
\centering
n_s&=1-2\frac{2\epsilon_1+\epsilon_2+\epsilon_3}{1-\epsilon_1},&n_T&=-2\frac{\epsilon_1+\epsilon_4}{1-\epsilon_1},&r&=8|\epsilon_1|\sum_{\lambda=L,R}\left|\frac{1}{\kappa^2Q_t}\right|c_A,
\end{align}
where $c_A$ stands for the field propagation velocity defined as,
\begin{equation}
\centering
\label{soundwave}
c_A^2=\frac{f_{,X}}{f_{,X}+2Xf_{,XX}}.\,
\end{equation}
The sound wave velocity for k-essence models is reduced to the following form,
\begin{equation}
\centering
c_A^2=\frac{a_0X+\alpha mX^m}{a_0X+(2m-1)\alpha mX^m}.
\end{equation}
Considering the latest Planck observational data \cite{Planck:2018jri}, the spectral index of the primordial curvature perturbations takes the value $n_\mathcal{S}=0.9649\pm 0.0042 $
with $67\%$ C.L and the tensor-to-scalar-ratio r must be $r<0.064$ with $95\%$ C.L. Last but not least, the tensor spectral index has yet to be observed. Consequently, in the near future a positive tilted tensor spectral index may be observed by LISA or NANOGrav rectifying blue gravitational waves compatible with Einstein-Chern-Simons gravitational theory with exotic kinetic terms. At this stage we shall present the methodology to evaluate the observational
indices during the first horizon crossing.
A convenient way to evaluate the observational indices is to use the initial and the final values of the scalar field instead of use wavenumbers. First of all, the evaluation of the final value of the scalar field is necessary. The final value of the scalar field can be derived by equating the first slow-roll index
$\epsilon_1$ in equation (\ref{e1k}) to unity. As a result, the initial value can be evaluated from the $e$-folding number, which defined as
$N=\int_{t_i}^{t_f}{Hdt}=\int_{\phi_i}^{\phi_f}{\frac{H}{\dot\phi}d\phi}$,
where the difference $t_f-t_i$ signifies the duration of the
inflationary era. Recalling the definition of $\dot\phi$ in Eq.
(\ref{fdot}), one finds that the proper relation from which the
initial value of the scalar field can be derived is,
\begin{equation}
\centering
\label{efolds1}
N=-\int_{\phi_i}^{\phi_f}\frac{3H^2\alpha_0}{V'}d\phi.\,
\end{equation}
As it stands, the present phenomenology is only valid for the case of $\alpha_0=\pm1$ and not for the purely noncanonical case of $f(R,\phi,X)\sim-a X^m$ corresponding to $\alpha_0=0$. In such a scenario the expression of $\dot\phi$ and in consequence the e-folding number are altered. Such model, although plausible, is not studied in the present article. Note also that an extra non minimal coupling between the scalar field and the Ricci scalar should alter the overall phenomenology as well. Although it is not a difficult task to study such model, the overall number of degrees of freedom increase therefore we refrain from including such extra coupling between the scalar field and curvature.  

In the following subsections two toy models are presented in order to ascertain the compatibility of such theory. As it turns out, the Chern-Simons scalar coupling function $\nu(\phi)$ can be chosen such that a blue tilted tensor spectral index can indeed be generated. Essentially this is the main difference between the pure k-essence model and the augmented by the Chern-Simons term. Let us now proceed with the subsequent models.

\subsection{Model with quadratic scalar potential}
The first proposed model involves a  slow rolling scalar field in a potential with quadratic form,
\begin{equation}
\centering
\label{pot}
V(\phi)=\lambda\left(\frac{\phi}{\kappa}\right)^2.
\end{equation}
The Chern-Simons scalar coupling function is considered to be an exponential function of the form,
\begin{equation}
\centering
\nu(\phi)=e^{-\gamma \kappa \phi},
\end{equation}
where $\gamma$ is a dimensionless parameter.
The slow-roll parameters of the model are given by the following expressions,
\begin{equation}
\centering
\epsilon_1\simeq \frac{2\alpha_0^3 \kappa^4-2.6  \alpha \lambda}{\alpha_0^4 \kappa^8 \phi^2},
\end{equation}
\begin{equation}
\centering
\epsilon_2\simeq \frac{2\kappa^4 \alpha_0^3(1-\alpha_0)-2.6\alpha \lambda }{\alpha_0^4 \kappa^8 \phi^2},
\end{equation}
\begin{equation}
\centering
\epsilon_3=0,
\end{equation}
\begin{equation}
\centering
\epsilon_4\simeq \frac{32 \gamma ^2 \lambda^2 (\gamma  \kappa  \phi -1)}{16 \alpha_0  \lambda^2 \gamma ^2 \kappa ^2  \phi ^2-9 \alpha_0^3 e^{2 \gamma  \kappa  \phi }}.
\end{equation}
The value of the k-essence field at the end of the inflationary era is derived setting the first slow-roll index into unity and therefore it reads,
\begin{equation}
\centering
\label{phif}
\phi_f\simeq\frac{\sqrt{2\alpha_0^3\kappa^4-2.6\alpha \lambda}}{\alpha_0^2\kappa^3}.
\end{equation}
Considering the form of the e-folding number in Eq.(\ref{efolds1}), the initial value of the scalar field is extracted and subsequently
the observed quantities. The initial value reads,
\begin{equation}
\centering
\label{phii}
\phi_i\simeq \frac{\sqrt{2\alpha_0^3\kappa^4(1+2N)-2.6\alpha \lambda}}{\alpha_0^{1/2}\kappa}.
\end{equation}
\begin{figure}[t!]
\centering
\label{plot1}
\includegraphics[width=17pc]{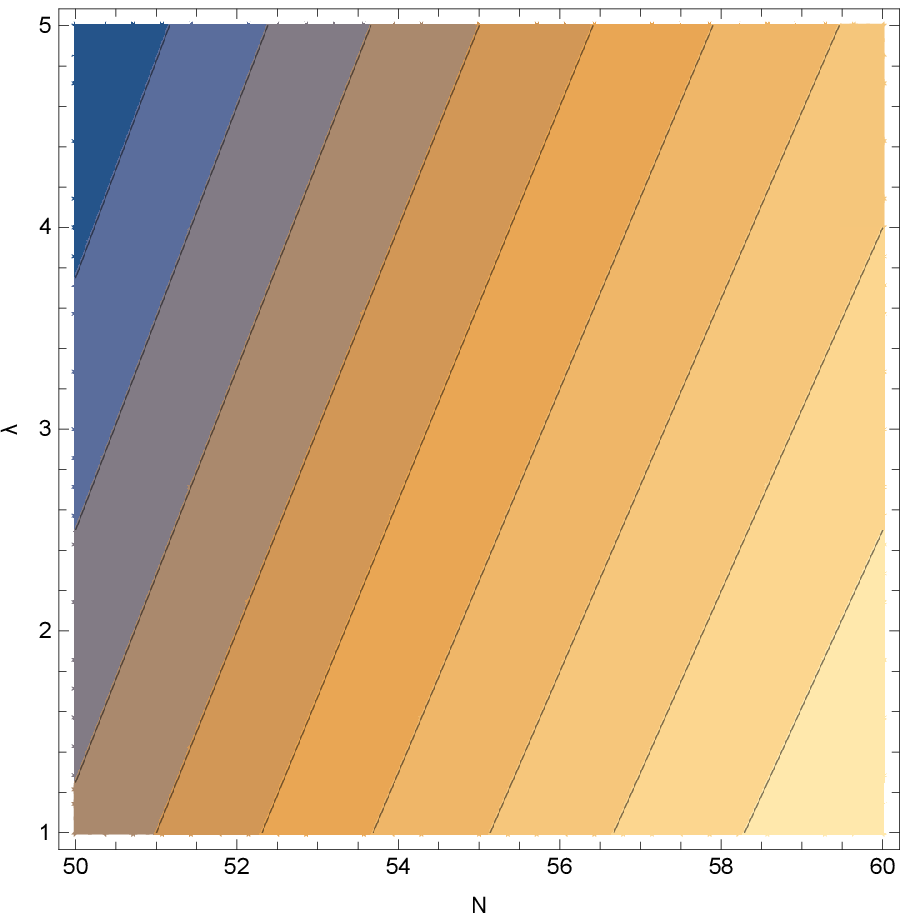}
\includegraphics[width=3pc]{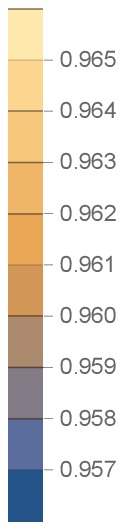}
\includegraphics[width=17pc]{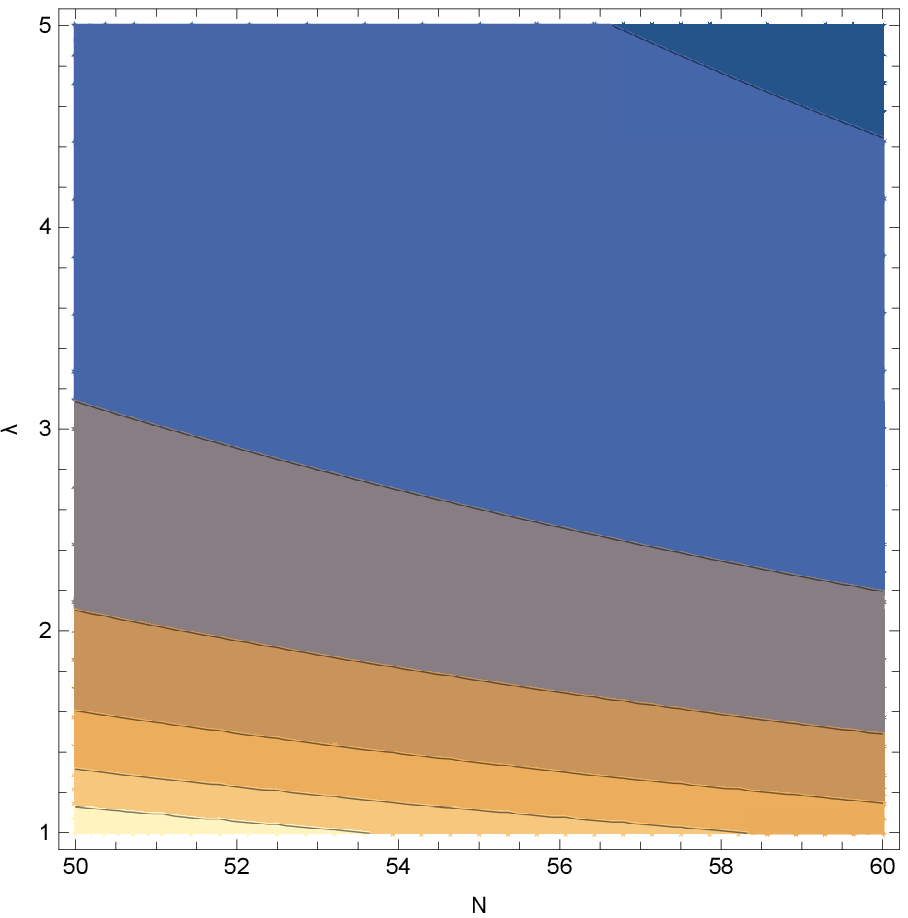}
\includegraphics[width=2.65pc]{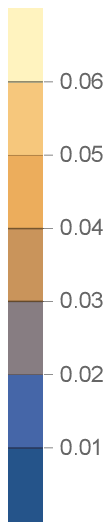}
\caption{Contour plots of the spectral index of primordial
curvature perturbations (left) and the tensor-to-scalar ratio
(right) depending on parameters $N$ and $\lambda$ ranging from
[50,60] and [1,5] respectively. } \label{plot1}
\end{figure}
\begin{figure}[t!]
\centering
\label{plot2}
\includegraphics[width=20pc]{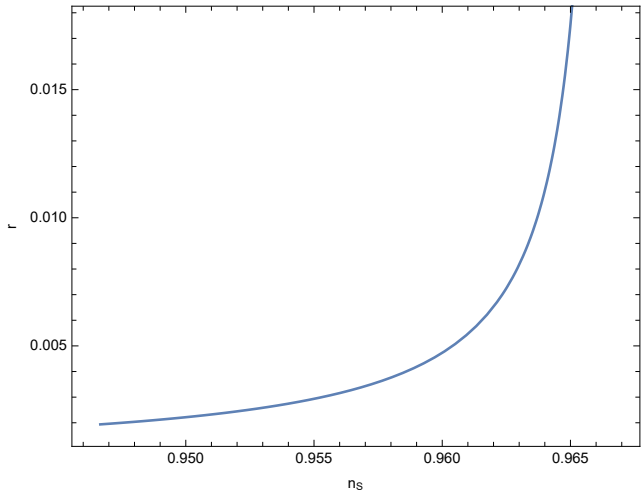}
\caption{Parametric plot of the spectral index of primordial
curvature perturbations as function of tensor-to-scalar ratio
 for the model with quadratic scalar potential for $N=60$ e-foldings. The parameter $\lambda$ for the indices is ranging from [0.1,30]. } \label{plot2}
\end{figure}

In order to ascertain compatibility with the latest observational data, in reduced Planck Units, the free parameters take the following values (N, $\gamma$, $\lambda$, $\alpha$, $\alpha_0$)=(60, -0.06, 1, -0.01, 1). As shown, the model is in good agreement with the experimental data \cite{Planck:2018jri} since, $n_\mathcal{S}=0.966$ and $r=0.0470219$. Furthermore, the tensor spectral
index takes the value $n_{\mathcal{T}}=0.018908$. The initial and the final numerical values of the scalar field are respectively $\phi_i=15.5572$ and $\phi_f=1.42361$ which means that the field decreases as time flows. The numerical values of the slow-roll indices are $\epsilon_1 =0.00837373$, $\epsilon_2=0.000110181$, $\epsilon_3=0$ and $\epsilon_4=-0.0177486$. Also, the model is free of ghosts since $c_A \simeq 1$. In Fig.\ref{plot1} the behaviour of the observed indices $n_{\mathcal{S}}$ and $r$ with respect to free parameters $N$ and $\lambda$ is shown from which it becomes abundantly clear that compatibility with Planck data can be achieved for a variety of values. Moreover, in Fig.\ref{plot2} the parametric plot between these two observed indices is depicted for a fixed e-folding number and running potential depth $\lambda$, from which one can easily extract the one to one correspondence between such parameters.

As expressed before, the tensor spectral index for such model turns out to be a positive number. This is a possible outcome for the k-essence model only if one admits the existence of a Chern-Simons coupling. In turn, this result can be connected to an enhancement in the intensity of the gravitational waves compared to the Einstein model, hence the reason such model is studied in the inflationary era, however to ascertain this a proper analysis of the late-time era for such scalar field assisted gravity is needed. Note also that due to the k-essence part, the field propagation velocity satisfies the relation $c_A\le1$ in order to respect causality and obtain a ghost free theory. Therefore the tensor to scalar ratio $r$ is always lesser, or at most equal to the expected result for the canonical scalar field. This can be used in order to effectively decrease the value of the tensor to scalar ratio if needed.

Finally, we
examine each approximation which was made in order to derive the previous results holds true. According to the previous set of parameters in reduced Planck units always, during the first horizon crossing, $\dot H \sim \mathcal{O}(10^{-1})$ and $H^2 \sim \mathcal{O}(10^{2})$ so the slow-roll assumption holds true. In addition, the kinetic term $ a_0X+\alpha mX^m  \sim \mathcal{O}(10^{-1})$ is negligible compared with the scalar potential $V \sim \mathcal{O}(10^{3})$ and lastly, $\ddot \phi=0$ and $3 H \dot \phi \sim \mathcal{O}(10^{2})$. Hence, the slow-roll conditions are valid.

\subsection{Model with trigonometric potential}
Suppose now that the potential of the scalar field is defined as,
\begin{equation}
V(\phi)=sin\left(\frac{\phi}{\phi_0}\right),
\end{equation}
where $\phi_0$ is a parameter of the model with $[\phi_0]=$ev and the potential amplitude was assumed to be equal to unity for simplicity. Note that the proper dimensions run with eV$^4$ for consistency. The Chern-Simons scalar coupling function is described by a simple power-law form,
\begin{equation}
\nu(\phi)=(\kappa \phi)^2,
\end{equation}
where again the prefactor was fixed to unity in order to avoid the appearance of additional irrelevant free parameters.

The first two gravitational equations for this model are,
\begin{equation}
\centering
H^2 \simeq \frac{\kappa^2 sin(\frac{\phi}{\phi_0})}{3},
\end{equation}
\begin{equation}
\centering
\dot H \simeq -\frac{cos(\frac{\phi}{\phi_0})cot(\frac{\phi}{\phi_0})}{7\alpha_0 \phi_0^2}.
\end{equation}
Let us now proceed with the overall phenomenology.
The slow-roll parameters of the model are given by the following expressions,
\begin{equation}
\centering
\epsilon_1\simeq \frac{cot(\frac{\phi}{\phi_0})^2}{2\alpha_0 \kappa^2 \phi_0^2},
\end{equation}
\begin{equation}
\centering
\epsilon_2\simeq \frac{2\alpha_0+cot(\frac{\phi}{\phi_0})^2}{2\phi_0^2 \alpha_0},
\end{equation}
\begin{equation}
\centering
\epsilon_3\simeq -\frac{\frac{\alpha}{2}cot(\frac{\phi}{\phi_0})(cos(\frac{\phi}{\phi_0})+cot(\frac{\phi}{\phi_0})csc(\frac{\phi}{\phi_0}))}{\alpha_0^4\kappa^4\phi_0^4-\alpha_0\alpha \kappa^2 \phi_0^2 cos(\frac{\phi}{\phi_0})cot(\frac{\phi}{\phi_0})}   ,
\end{equation}
\begin{equation}
\centering
\epsilon_4\simeq \frac{16\kappa^6\phi cos(\frac{\phi}{\phi_0})^2(\phi-\phi_0 cot(\frac{\phi}{\phi_0}))}{16\alpha_0\kappa^8\phi^2\phi_0^2cos(\frac{\phi}{\phi_0})^2-9\alpha_0^3\phi_0^4} .
\end{equation}
Similar to the previous model, the value of the scalar field at the end of inflation is derived from the equation
$\epsilon_1=1$ however, due to the perplexed form of the equations they are omitted. A real value can be obtained in order to avoid complex valued scalar field, something which becomes apparent subsequently. Also, by making use of the definition of the e-folding number the initial value of the scalar field during the first horizon crossing is extracted. Obviously it depends of $N$ and $\phi_f$ however as in the case of $\phi_f$ it was deemed lengthy and thus unnecessary to be shown. Let us proceed with the numerical results and comparison with observations.

\begin{figure}[t!]
\centering
\label{plot3}
\includegraphics[width=17pc]{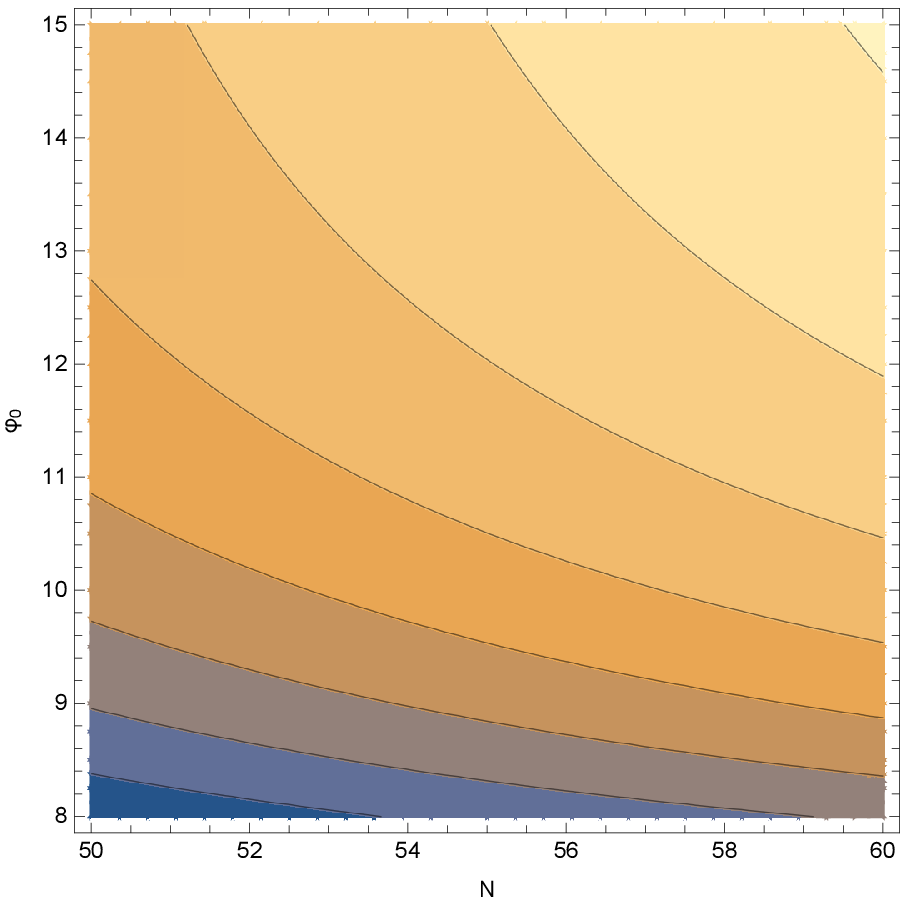}
\includegraphics[width=3pc]{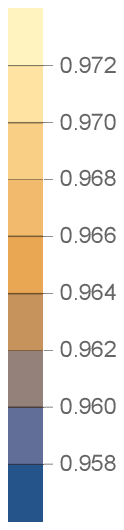}
\includegraphics[width=17pc]{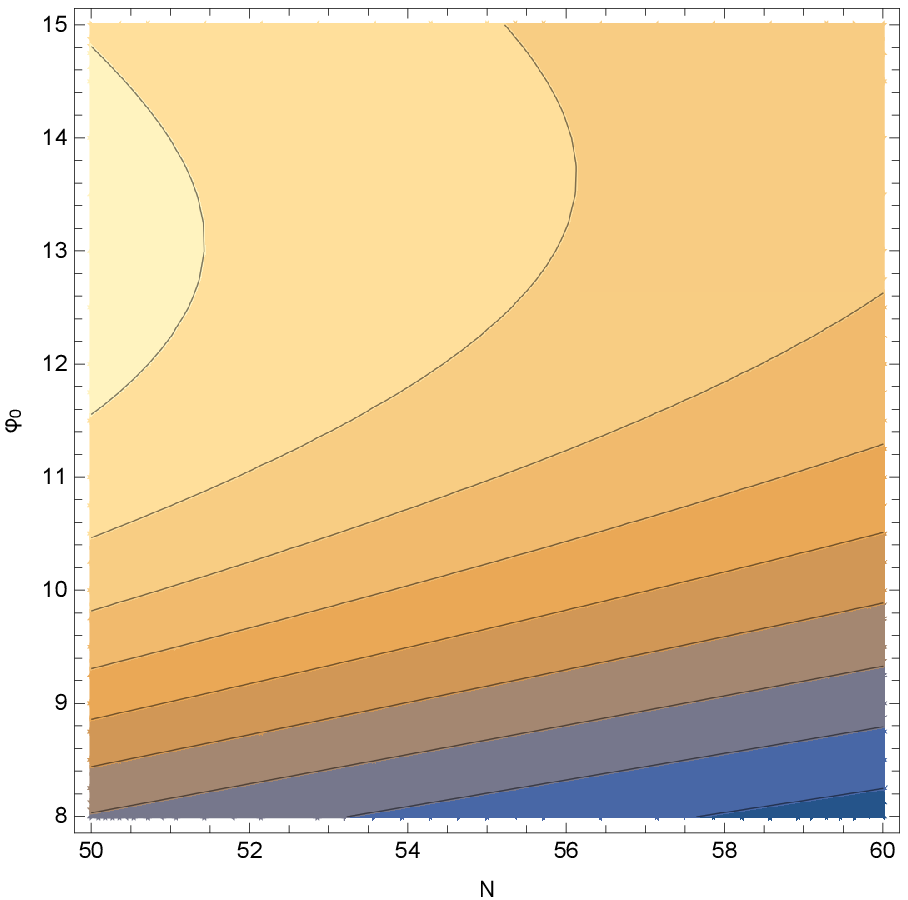}
\includegraphics[width=2.65pc]{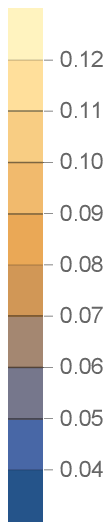}
\caption{Contour plots of the spectral index of primordial
curvature perturbations (left) and the tensor-to-scalar ratio
(right) depending on parameters $N$ and $\phi_0$ ranging from
[50,60] and [8,15] respectively. } \label{plot3}
\end{figure}
Specifying the free parameters of the theory one could produce results compatible with the observational values for the
spectral indices and the tensor-to-scalar ratio introduced in Eq. (\ref{observed}). Assuming that (N, $\alpha_0$, $\alpha$, $\phi_0$)=(60, 1, -0.01, 8.9), in reduced Planck units, so for $\kappa^2$=1, the model at hand produces acceptable results, since $n_\mathcal{S}=0.964101$ and $r=0.0519585$ are both compatible with observations. Furthermore, the tensor spectral
index takes the value $n_{\mathcal{T}}=0.00736323$. The blue shifted tensor spectral index is again generated from the presence of 
Chern-Simons term. The initial and the final numerical values of the scalar field are respectively $\phi_i=9.65185$ and $\phi_f=-0.705439$ which means that the field is decreasing with time. The numerical values of the slow-roll indices are $\epsilon_1=0.00176425$, $\epsilon_2=0.0143889$, $\epsilon_3\sim  \mathcal{O}(10^{-7})$ and $\epsilon_4 \simeq -0.00543937$. Also, the sound wave velocity is almost equal to unity and real, thus the model  respects causality and is also free of ghosts respectively. The aforementioned designation of free parameters is obviously not the only one capable of producing compatible with the observations results. In figures \ref{plot3} and \ref{plot4} we present the respective contour plots of the scalar spectral index of primordial primordial curvature perturbations and the tensor to scalar ratio and also the parametric plot. Both observed indices are shown as functions of auxiliary parameters $N$ and $\phi_0$ for the first while in the latter case it becomes clear that allowed pairs of viable $n_S$ and $r$ reside in a slim area of viability.

At this stage it should be stated that the sign of the tensor spectral index is not always positive however, in both k-essence models which we present, the auxiliary parameters were in an essence fine tuned in order to obtain such result. Obviously the fact that $n_{\mathcal{T}}$ appears to be positive is due to the existence of the Chern-Simons scalar coupling function $\nu(\phi)$ and appears whenever the condition $\epsilon_4+\epsilon_1<0$ is satisfied however to our knowledge there exists no universal behaviour that once satisfied, it guarantees that a blue tilted tensor spectral index shall be produced. In general, such result is entirely model dependent and therefore extra care is needed if one wishes to produce a positive tensor spectral index for the k-essence model since not only does one need compatible results with Planck data, but due to the existence of the non canonical kinetic term $\sim \alpha X^2$, a well behaved field propagation velocity is also needed. In both models, as demonstrated, such case is indeed possible.

Finally, let us discuss here and validate whether the approximations assumed in the section II hold true. First of all, we shall check the validity of the slow-roll approximations, $\dot H \sim \mathcal{O}(10^{-4})$ and $H^2\sim \mathcal{O}(10^{1})$ hence the approximation, $\dot H\ll H^2$ holds true. In addition the kinetic term of the k-essence scalar field is $a_0 X+\alpha mX^m \sim \mathcal{O}(10^{-4})$ while the scalar potential is  $V\sim\mathcal{O}(10^{-1})$, hence the condition $a_0 X+\alpha mX^m\ll V$ is valid. Lastly, comparing the term $\ddot \phi\sim \mathcal{O}(10^{-3})$ with the term $H \dot \phi\sim \mathcal{O}(10^{-2})$, it is clear that the condition $\ddot \phi\ll 3H\dot \phi$ is satisfied. Thus, all the approximations are valid.

\begin{figure}[t!]
\centering
\label{plot4}
\includegraphics[width=17pc]{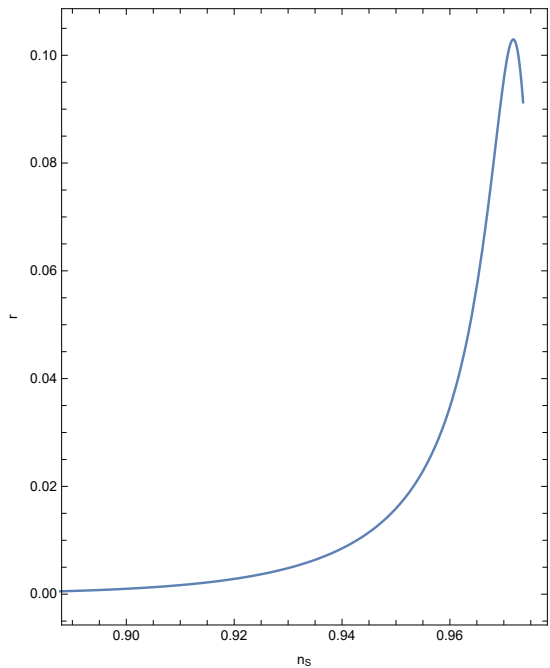}
\caption{Parametric plot of the spectral index of primordial
curvature perturbations as function of tensor-to-scalar ratio
 for the model with quadratic scalar potential for $N=60$ e-foldings. The parameter $\phi_0$ for the indices is ranging from [3,20]. } \label{plot4}
\end{figure}

Therefore, both examined models serve as examples of viable k-essence model that admit blue tilted tensor spectral index. Although model dependent, such result when it manifests can lead to interesting results in subsequent cosmological eras.

\newpage
\section{THEORETICAL FRAMEWORK OF INFLATIONARY DIRAC-BORN-INFELD MODELS IN EINSTEIN-CHERN-SIMONS GRAVITY}
In this section we shall investigate the inflationary phenomenology of Dirac-Born-Infeld string-inspired models in Einstein-Chern-Simons gravity examining their viability. The gravitational action which is utilized in order to extract all the information about the primordial Universe is Eq.(\ref{action}). As k-essence models, Dirac-Born-Infeld models belong to a general class of $f(R, \phi, X)$ theories. Hereafter, the general form of the
$f(R, \phi, X)$ function is assumed to be the following,
\begin{equation}
\centering
\label{f2}
f(R, \phi, X)=\frac{R}{\kappa^2}-2K(\phi)^{-1}\sqrt{1+2K(\phi)X(\phi)}-2V(\phi),
\end{equation}
where once again R is the Ricci scalar, $K(\phi)$ is a function which depends from the scalar field and $V(\phi)$ is the scalar potential. It should be stated that the auxiliary function $K(\phi)$ must have mass dimensions of eV$^{-4}$ for the sake of consistency.
Substituting Eq.(\ref{f2}) into the equations of motion (\ref{motion1}-\ref{motion3}) the following equations arise,
\begin{equation}
\centering
\label{eq1dbi}
\frac{3H^2}{\kappa^2}=\frac{VK+\frac{1}{\sqrt{1+2KX}}}{K},
\end{equation}
\begin{equation}
\centering
\label{eq2dbi}
\frac{\dot H}{\kappa^2}=\frac{X}{\sqrt{1+2KX}},
\end{equation}
\begin{equation}
\centering
\label{eqdbi}
\frac{1}{\sqrt{1+2KX}}[\ddot \phi+3H\dot \phi+\frac{K'(1+KX)}{K^2}-\dot \phi(\frac{K'\dot \phi X+K \dot X}{1+2KX})]+V'=0.
\end{equation}
The above equations seem quite perplexed to be solved analytically. In order to proceed our analysis further approximations are required.
First of all, the factor $1/\sqrt{1+2KX}$ is expanded via a Taylor series by making the assumption that $KX\ll1$. In turn, the linear term which emerges is subleading compared to the scalar potential if one admits the slow-roll conditions are valid. During the inflationary era the slow-roll conditions are imposed as presented,
\begin{align}
\label{approx}
\centering
\dot H&\ll H^2,& X&\ll V,& \ddot\phi\ll3 H\dot\phi.\
\end{align}
As the contribution of the kinetic term X is minor, the products $KX$, $K'\dot \phi X$ and $K\dot X$ can be discarded from the above equations. Considering all the approximations, 
the equations of motion are simplified further,
\begin{equation}
\centering
\label{eq1dbifinal}
H^2 \simeq \kappa^2 \frac{VK+1}{K},
\end{equation}
\begin{equation}
\centering
\label{eq2dbifinal}
\dot H\simeq \kappa^2 X,
\end{equation}
\begin{equation}
\centering
\label{eq3dbifinal}
3H\dot \phi +\frac{K'}{K^2}+V'\simeq 0.
\end{equation}
Solving Eq.(\ref{eq3dbifinal}) one can get easily,
\begin{equation}
\centering
\label{fdotdbi}
\dot \phi \simeq -\frac{1}{3H}\left(V+\frac{K'}{K^2}\right).
\end{equation}
As mentioned in k-essence case, the dynamics of Dirac-Born-Infeld inflation in the context of $f(R, \phi, X)$ gravity can be described in terms of the slow-roll parameters \cite{Hwang:2005hb}.
The first slow-roll index is given by the expression,
\begin{equation}
\centering
\label{e1dbi}
\epsilon_1 \simeq \frac{(3K'+K(1+KV)V')^2}{18\kappa^2(1+KV)^2},
\end{equation}
while the rest indices are given by quite perplexed equations. Last but not least, parameter E of the model is described by the following equation,
\begin{equation}
E=\frac{81 \sqrt{3}}{\left(\frac{K \left(K \left(27-(K V+1) V'^2\right)-6 K' V'\right)-\frac{9 K'^2}{K V+1}}{K^2}\right)^{3/2}}.
\end{equation}

\section{Compatibility OF INFLATIONARY DIRAC-BORN-INFELD MODELS WITH OBSERVATIONS}
Compatibility with the latest Planck data is discussed extensively in section III. For Dirac-Born-Infeld models the equations of the observed indices  (\ref{observed}) remain the same. The sound wave velocity $c_A^2$ is given by the equation,
\begin{equation}
\label{dbica}
\centering
c_A^2=1-\frac{\left(3 K'+K (K V+1) V'\right)^2}{27K^2 (K V+1)},
\end{equation}
while the number of e-foldings is:
\begin{equation}
\centering
\label{dbiefoldings}
N=-\int_{\phi_i}^{\phi_f}  \frac{3 K (K V+1)}{3 K'+K (K V+1) V'}
d\phi.\,
\end{equation}
As shown, the field propagation velocity remains strictly lesser than unity if the condition $KV+1>0$ is satisfied. In this case it is expected that the model is free of ghost instabilities while simultaneously respects causality. Therefore both cases are described by the same condition. On the other hand the e-folding number seems to be depending on both noncanonical function $K(\phi)$ but also the scalar potential $V(\phi)$, along with their derivatives. This suggests that the overall phenomenology is expected to be quite perplexed. For that reason we shall limit our work to a single trivial example of constant scalar potential to simplify the expression of the e-folding number. The numerical results are available in the following subsection.

\subsection{Model with Dirac-Born-Infeld scalar field and constant potential}
It is presented inflationary phenomenology of Dirac-Born-Infeld scalar field with Chern-Simons string-corrections. The scalar potential is considered to be constant,
\begin{equation}
\centering
\label{dbipot}
V(\phi)=\Lambda,
\end{equation}
while the Chern-Simons scalar coupling function $\nu(\phi)$ and the function K are given by the exponential and power-law form respectively,
\begin{equation}
\centering
\label{cs}
\nu(\phi)=e^{\kappa \phi},
\end{equation}
\begin{equation}
\centering
\label{F}
K(\phi)=\kappa^6\phi^{2}.
\end{equation}
Note that while $\nu(\phi)$ is indeed dimensionless, the scalar potential, even though constant, must have mass dimensions of eV$^4$ therefore $[\Lambda]=$ev$^4$. Also, hereafter natural units are used for simplification therefore the replacement $\kappa=1=M_P$ is used.

The first two gravitational equations for the model are,
\begin{equation}
\centering
H^2\simeq \frac{1}{3} \left(\Lambda +\frac{1}{\phi ^2}\right),
\end{equation}
\begin{equation}
\centering
\dot H \simeq-\frac{2}{3 \left(\Lambda  \phi ^6+\phi ^4\right)}.
\end{equation}

The dynamics of the model is described by the following slow-roll parameters,
\begin{equation}
\centering
\epsilon_1\simeq \frac{2}{\left(\Lambda \phi ^3+\phi \right)^2},
\end{equation}
\begin{equation}
\centering
\epsilon_2\simeq \frac{6 \Lambda  \phi ^2+4}{\left(\Lambda  \phi ^3+\phi \right)^2} ,
\end{equation}
\begin{equation}
\centering
\epsilon_3 \simeq \frac{24 \Lambda  \phi ^2+12}{\left(\Lambda  \phi ^3+\phi \right)^2 \left(3 \Lambda  \phi ^4+3 \phi ^2-4\right)},
\end{equation}
\begin{equation}
\centering
\epsilon_4\simeq \frac{e^{2\phi } (6 -2 \phi )}{\phi ^2 \left( e^{2\phi }- \phi ^6/2\right) \left(\Lambda  \phi^2+1\right)} .
\end{equation}
Equating to unity the
first slow-roll index and utilizing the e-folding number $N$ in Eq. (\ref{dbiefoldings}), one obtains the following values for the scalar field,
\begin{equation}
\centering
\label{phif}
\phi_f \simeq \frac{\sqrt{\frac{\left(\Lambda -\sqrt[3]{27 \Lambda ^4+\Lambda ^3+3 \sqrt{3} \sqrt{\Lambda ^7 (27 \Lambda +2)}}\right)^2}{\Lambda ^2 \sqrt[3]{27 \Lambda ^4+\Lambda ^3+3 \sqrt{3} \sqrt{\Lambda ^7 (27 \Lambda +2)}}}}}{\sqrt{3}},
\end{equation}
\begin{equation}
\centering
\label{phii}
\phi_i\simeq \frac{\sqrt{\frac{3 \sqrt{\frac{\left(\Lambda  \sqrt[3]{27 \Lambda ^4+\Lambda ^3+3 \sqrt{3} \sqrt{\Lambda ^7 (27 \Lambda +2)}}+\left(27 \Lambda ^4+\Lambda ^3+3 \sqrt{3} \sqrt{\Lambda ^7 (27 \Lambda +2)}\right)^{2/3}+\Lambda ^2\right)^2}{9 \Lambda ^2 \left(27 \Lambda ^4+\Lambda ^3+3 \sqrt{3} \sqrt{\Lambda ^7 (27 \Lambda +2)}\right)^{2/3}}+8 \Lambda N}-3}{\Lambda }}}{\sqrt{3}}.
\end{equation}

\begin{figure}[t!]
\centering
\label{plot5}
\includegraphics[width=17pc]{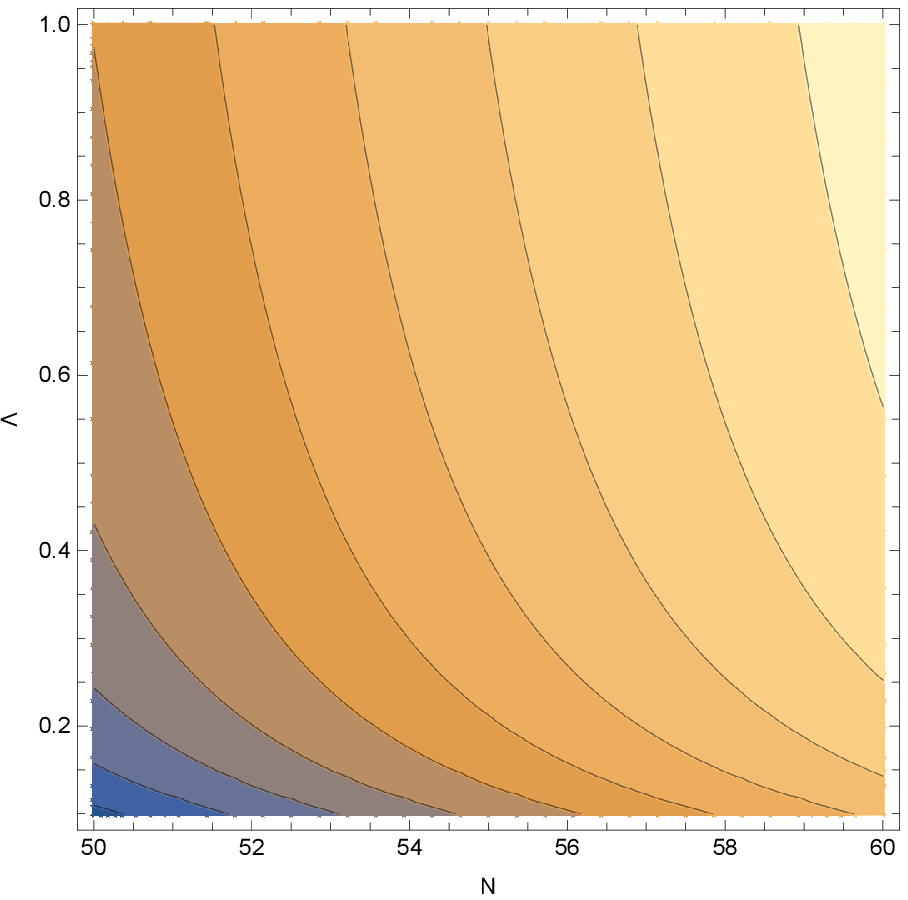}
\includegraphics[width=2.65pc]{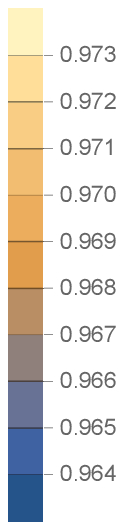}
\includegraphics[width=17pc]{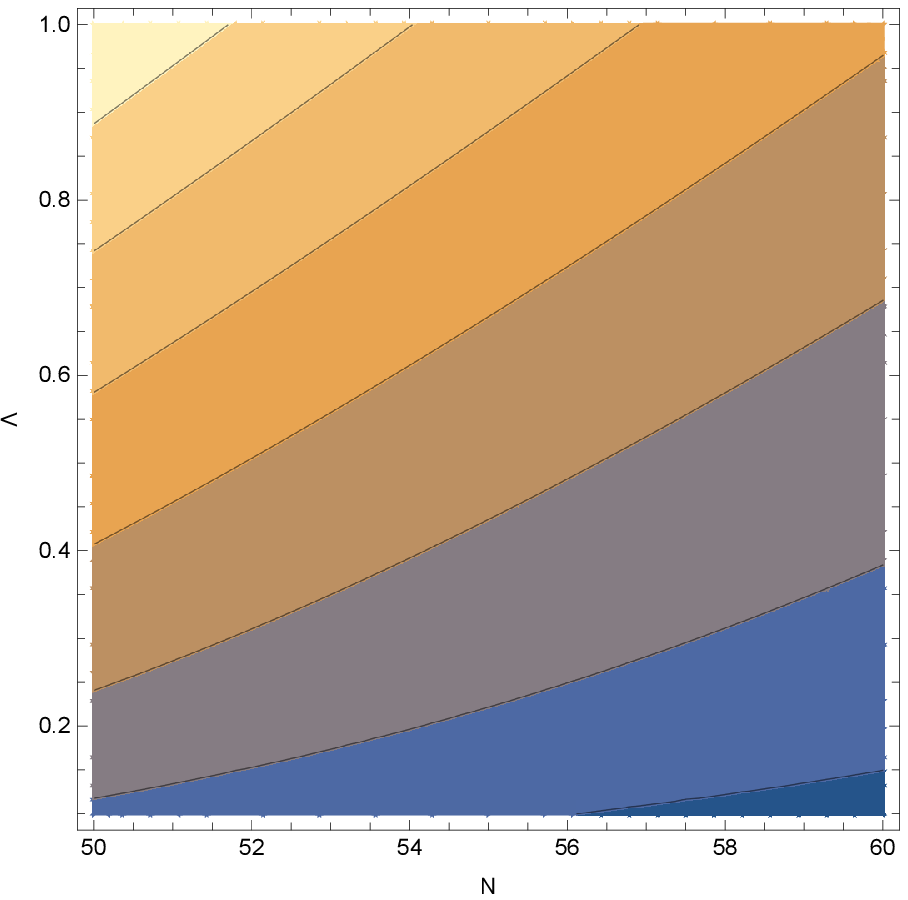}
\includegraphics[width=2.65pc]{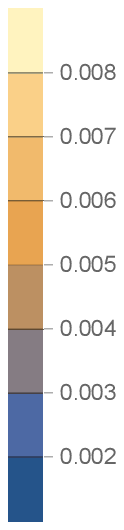}
\caption{Contour plots of the spectral index of primordial
curvature perturbations (left) and the tensor-to-scalar ratio
(right) depending on parameters $N$ and $\Lambda$ ranging from
[50,60] and [0.1,1] respectively. } \label{plot5}
\end{figure}
\begin{figure}[t!]
\centering
\label{plot6}
\includegraphics[width=20pc]{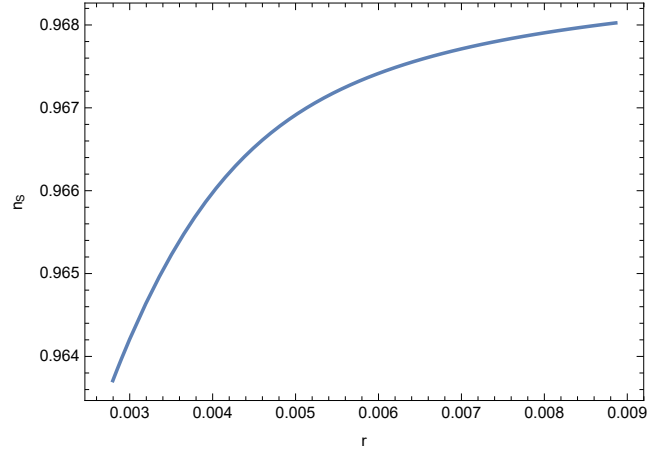}
\caption{Parametric plot of the spectral index of primordial
curvature perturbations as function of tensor-to-scalar ratio
 for the model with constant scalar potential for $N=60$ e-foldings. The value of the potential $\Lambda$ for the indices is ranging from [0.1,1]. } \label{plot6}
\end{figure}
The previous statement referring to difficulties in the overall phenomenology due to the presence of a noncanonical term $K$ stands corrected as for the trivial case of a constant scalar potential throughout the inflationary era, the expressions of the initial and final value of the scalar field as functions of the free parameters are quite complicated. This does not mean that other potentials are prohibited per se but rather $\phi$ dependent potentials manifest certain difficulties in the nature of the scalar field as complex numbers arise frequently. Let us proceed with the results of the model at hand. 

Numerically speaking, let us assign the following values
to the free parameters, always in reduced Planck units ($N$, $\Lambda$)=(50, 0.4) the observational indices take the values $n_{\mathcal{S}}=0.966886$, $n_{\mathcal{T}}=0.0351517$ and $r=0.00495707$ which are compatible with the latest observations. The scalar field seems to decrease with time since $\phi_i=5.41455$ and $\phi_f=1.00644$. Also, the model is free of ghosts since $c_A\simeq1$, or the latter term in Eq. (\ref{dbica}) is positive but quite subleading. Furthermore, the slow-roll indices have the following numerical values
$\epsilon_1=0.000421172$, $\epsilon_2 \simeq 0.0156595$, $\epsilon_3\simeq 0.0000554044$ and $\epsilon_4 \simeq -0.017997$, therefore the respective slow-roll conditions are expected to apply in this situation.

The above designation as shown leads to a quite negative slow-roll index $\epsilon_4$ thus the tensor spectral index changes sign and becomes positive. Therefore the fact that the tensor spectral index can become blue tilted is not limited only in the k-essence models but applies to various models, either canonical or noncanonical, provided that $\epsilon_4+\epsilon_1<0$. Therefore, viability of Chern-Simons string inspired noncanonical theories have interesting implications in general as the generation of a blue tilted tensor spectral index may be indicative of theories that predict enhanced intensity of gravitational waves. As mentioned before, this can be ascertained once a proper study of the late-time era is performed.

The last step is to check whether the approximations considered in section IV holds true. According to the previous set of parameters in reduced Planck units always, during the first horizon crossing, $\dot H \sim \mathcal{O}(10^{-1})$ and $H^2 \sim \mathcal{O}(10^{-5})$  so the slow-roll assumption holds true. In addition $X \sim \mathcal{O}(10^{-5})$ while $V \sim \mathcal{O}(10^{-1})$ and lastly, $\ddot \phi \sim \mathcal{O}(10^{-3})$ and $3 H \dot \phi \sim \mathcal{O}(10^{-2})$. Hence, the slow-roll conditions are valid. All that remains is to ascertain the validity of the rest approximations. It turns out that the term $K'KX\sim \mathcal{O}(10^{-4})$ is negligible compared to the term $K'\sim \mathcal{O}(10^{-1})$ thus, our approximation $K'KX\ll K'$ holds true. Finally the term $\dot \phi(\frac{K'\dot \phi X+K \dot X}{1+2KX})\sim \mathcal{O}(10^{-5})$ is quite smaller in order of magnitude than $3H\dot \phi$ concluding that all our approximations are indeed valid.

Overall, the Dirac-Born-Infeld model is capable of producing compatible with the Planck data results while simultaneously it manifests a blue tilted tensor spectral index provided that a Chern-Simons scalar coupling function is present.

\section{Conclusions}
In the present article the viability of certain noncanonical scalar theories of gravity in the presence of string corrective term was studied. Initially, the case of k-essence models was examined by introducing the proper phenomenology. By making use of the slow-roll conditions and simplifying in accordance the equations of motion we found that the inflationary era can manifest compatible with the observations results. The inclusion of a Chern-Simons scalar coupling function does not spoil compatibility with the Planck data however as showcased previously, it can lead to a blue tilted tensor spectral index $n_{\mathcal{T}}$ something which is impossible in the case of $\nu(\phi)=0$. Subsequently, a second noncanonical model and in particular a Dirac-Born-Infeld model was studied. Considering a trivial substitution on the scalar potential and admitting a constant value during the inflationary era, compatible results can once again be generated with the tensor spectral index being again positive. Although there exists no universal condition which guarantees a blue tilted tensor spectral index if it is satisfied, apart from the condition $\epsilon_4+\epsilon_1<0$, the results even though model dependent are in agreement with the latest Planck data. The fact that a blue tilted tensor spectral index can be manifested, although it has yet to be observed, makes a powerful prediction about the behaviour of the overall theory and also gravitational waves. A blue tilted tensor spectral index can be connected to an enhcanced signal in the gravitational waves relative to the expected one from GR which in principle could be observed in the near future by LISA or NANOGrav for instance. To ascertain whether such enhancement is feasible, a proper analysis of the late-time era is needed by keeping the same values of the free parameters that produce a viable inflationary era. Truthfuly the overall model may be altered by assuming for instance the existence of an $f(R)$ function which receives two contribution, a linear contribution as in GR which is dominant mainly in the inflationary era and an additional one which, although subleading in the early era, dominates the late time era. This is an interesting scenario for non canonical theories scalar-tensor theories in general, augmented by a Chern-Simons scalar coupling functions and we leave this study for a future work.

\end{document}